\documentclass[conference]{IEEEtran}

\IEEEoverridecommandlockouts

\usepackage[noadjust]{cite} % noadjust is a good option for tight layouts
\usepackage{amsmath,amssymb,amsfonts}
\usepackage{algorithmic}
\usepackage{graphicx}
\usepackage{textcomp}
\usepackage{xcolor}
\usepackage{colortbl} % For cell/column coloring
\usepackage{array}    % For advanced column specifications
\usepackage{url}
\usepackage{booktabs} % For better tables
\usepackage{multirow}
\usepackage{subcaption} % For subfigures, if needed
\usepackage[hidelinks]{hyperref} % For clickable links/references

\definecolor{asynccolumnbg}{gray}{0.9}
\def\BibTeX{{\rm B\kern-.05em{\sc i\kern-.025em b}\kern-.08em
    T\kern-.1667em\lower.7ex\hbox{E}\kern-.125emX}}

\title{AsyncVoice Agent: Real-Time Explanation \\for LLM Planning and Reasoning}

\author{
\IEEEauthorblockN{Yueqian Lin$^{1,2}$, Zhengmian Hu$^{2}$, Jayakumar Subramanian$^{2}$, Qinsi Wang$^{1,2}$\\Nikos Vlassis$^{2}$, Hai ``Helen'' Li$^{1}$, Yiran Chen$^{1}$}
\IEEEauthorblockA{$^{1}$Duke University, Durham, NC, USA\qquad$^{2}$Adobe Research, San Jose, CA, USA }
}

\begin{document}
\maketitle

%-------------------------------------------------------------------------
\begin{abstract}
Effective human-AI collaboration on complex reasoning tasks requires that users understand and interact with the model's process, not just receive an output. However, the monolithic text from methods like Chain-of-Thought (CoT) prevents this, as current interfaces lack real-time verbalization and robust user barge-in. We present AsyncVoice Agent, a system whose asynchronous architecture decouples a streaming LLM backend from a conversational voice frontend. This design allows narration and inference to run in parallel, empowering users to interrupt, query, and steer the model's reasoning process at any time. Objective benchmarks show this approach reduces interaction latency by more than 600$\times$ compared to monolithic baselines while ensuring high fidelity and competitive task accuracy. By enabling a two-way dialogue with a model's thought process, AsyncVoice Agent offers a new paradigm for building more effective, steerable, and trustworthy human-AI systems for high-stakes tasks.\footnote{This work was supported by NSF 2112562 and ARO W911NF-23-2-0224.}
\end{abstract}


%-------------------------------------------------------------------------
\section{Introduction}
\label{sec:intro}

Large Language Models (LLMs) increasingly tackle complex tasks by generating step-by-step reasoning, often in a ``Chain-of-Thought'' (CoT) format~\cite{wei2022chain}. While intended to provide transparency, these reasoning traces are typically delivered as monolithic, text-based outputs. This presentation format is particularly ill-suited for spoken language interfaces, as it forces users into a passive listening role for potentially long, uninterrupted monologues, hindering comprehension and preventing real-time interaction~\cite{zhang2024verbosity,lin2025speechprune,lin2025voice}. This fundamental mismatch between the static output of reasoning models and the dynamic nature of human conversation limits the potential for fluid, collaborative human-agent partnerships.

Consequently, enhancing the interpretability of these reasoning chains has become a significant focus of ongoing research. While raw CoT aims for transparency, its verbosity can be a significant barrier~\cite{zhang2024verbosity}. Some approaches use text-based summarization or structuring techniques to offer post-hoc digestibility~\cite{zhu2025iterative}, and dialogue systems have been proposed to improve explanations~\cite{ho2025dialogue}. However, these typically do not operate on the live, streaming internal state of a separate reasoning LLM. More recent work has explored asynchronous AI agents for real-time tool use with voice interaction, focusing on the agent's task execution (e.g., booking a flight while conversing)~\cite{ginart2024asynchronous}. While this addresses the agent's external actions, the challenge of providing real-time, interactive explanations of an agent's \emph{internal reasoning stream}, and allowing for user interruption during this process, remains a significant, unaddressed gap.

To fill this gap, this paper presents the architecture and demonstration of \textbf{AsyncVoice Agent}, a system designed to serve as an interactive, real-time voice interface for explaining an LLM's ongoing reasoning processes. Rather than verbalizing a static block of text, our system narrates each thought process as it streams from the model. The primary contribution is a system that enables a truly asynchronous dialogue about a live reasoning process, allowing a user to seamlessly interrupt the agent's explanation to ask questions or provide feedback. This capability transforms the interaction from a passive user experience into a collaborative dialogue, filling a crucial gap for explainability and trust. This paper details the system's architecture and its novel components.

%-------------------------------------------------------------------------
\section{AsyncVoice Agent Architecture}
\label{sec:method}

\begin{figure*}[t]
    \centering
    \includegraphics[width=0.9\linewidth]{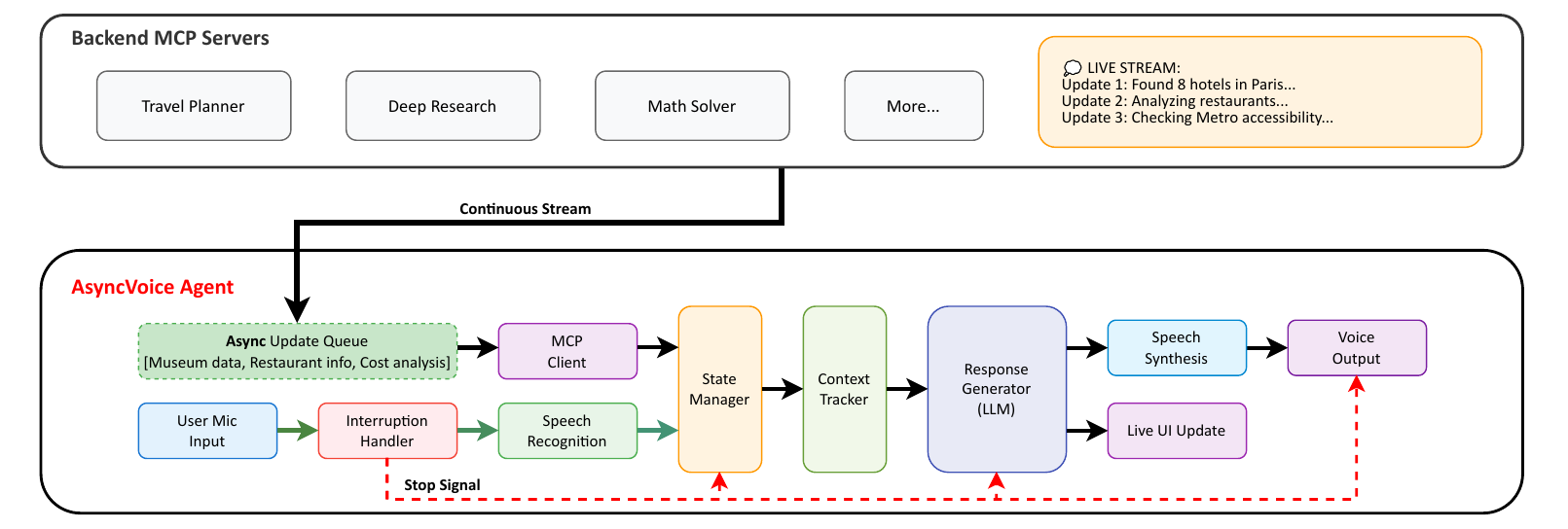}
    \caption{The high-level architecture of the AsyncVoice Agent. The system features two primary components: the \textbf{Backend MCP Servers} that generate a \texttt{Continuous Stream} of reasoning steps, and the \textbf{AsyncVoice Agent} pipeline, which verbalizes these steps and manages asynchronous user interruptions via a \texttt{Stop Signal}.}
    \label{fig:architecture}
\end{figure*}

The AsyncVoice Agent is a real-time voice interface system designed to provide interactive explanation of an LLM's reasoning process. Our implementation builds upon the foundational \texttt{RealtimeVoiceChat} project\footnote{\url{https://github.com/KoljaB/RealtimeVoiceChat}}, adapting its core real-time ASR and WebSocket communication layer. Our primary contributions are the design and integration of several novel components that enable the explanation of an external LLM's reasoning stream. As illustrated in Figure~\ref{fig:architecture}, our architecture adds new backend reasoning modules and modifies the core agent pipeline. The complete system comprises three core subsystems: (1) the adapted WebSocket layer for bidirectional audio streaming, (2) our novel modular Model Context Protocol (MCP) servers for specialized reasoning tasks, and (3) a multi-threaded speech processing pipeline featuring Azure TTS integration for low-latency voice interaction.

\subsection{WebSocket Communication Infrastructure}
The system employs a FastAPI-based WebSocket server to handle concurrent bidirectional data streams. The protocol supports two distinct message types: JSON-formatted text messages for control commands and transcription updates, and binary messages containing raw PCM audio data. This dual-channel approach enables sub-second latency for voice interactions while maintaining reliable command transmission. To manage connection-specific conversation context and prevent resource leaks, the server uses a dedicated context manager that tracks active connections. Crucially, this handler maintains the conversational state by chronologically integrating all events, such as streamed backend reasoning steps and user voice interruptions, into a single, unified session context. This design also supports potential multi-user and multi-agent collaborative sessions.

\subsection{MCP-Based Modular Reasoning Architecture}
The backend reasoning capabilities are provided by the specialized \textbf{Backend MCP Servers} shown in Figure~\ref{fig:architecture}. These servers are highly modular, allowing different underlying large language models to be configured for specific tasks. For instance, the \texttt{Travel Planner} is configured to use a specialized reasoning model for its complex planning capabilities, while the \texttt{Math Solver} leverages the precision of GPT-4o \cite{hurst2024gpt}. Regardless of the reasoning model used, each server adheres to the standardized MCP. They emit progress updates and final answers using a consistent notification format (\texttt{ctx.notification}). The agent's \texttt{MCP Client} subscribes to this stream and processes updates based on semantic prefixes: \texttt{Thinking:} for an intermediate reasoning step, \texttt{Content:} for a status update, and \texttt{Answer:} for the final response. This protocol standardization allows the AsyncVoice Agent to adapt to any MCP-compliant backend with minimal effort. A final \texttt{COMPLETE} signal is used to trigger UI state transitions on the front-end and update the state manager accordingly.

\subsection{Multi-Threaded Speech Processing Pipeline}
To minimize response latency, the core of the agent orchestrates four concurrent processing threads. This pipeline manages the data flow from the \texttt{Response Generator (LLM)} to the final \texttt{Voice Output} (as shown in Figure~\ref{fig:architecture}). A Request Processing Thread handles incoming user queries. An LLM Inference Thread processes text through the agent's \textbf{explainer LLM} (powered by GPT-4o) to formulate natural-language explanations of the backend's reasoning or to respond directly to user questions. For audio generation, our system integrates Azure TTS in a dual-stage process: a \textit{Quick Thread} synthesizes the first part of a response for immediate playback, and a \textit{Final Thread} generates the complete audio. A crossfade mechanism ensures a seamless transition, creating continuous, natural-sounding speech output.

\subsection{Turn Detection and Interruption Handling}
The system implements turn detection using a DistilBERT-based transformer that performs binary sentence completion classification. The model processes text segments with a maximum sequence length of 128 tokens and outputs probability scores indicating whether a sentence is complete or incomplete. These probability scores are mapped to dynamic pause durations through configurable anchor points, enabling natural conversation flow based on semantic context.

The interruption handling mechanism operates across multiple system layers to ensure responsive interaction. Audio worklet processors in the browser continuously monitor microphone input and can trigger interruption signals with sub-100ms latency. Upon detecting user speech, the system immediately issues a stop signal that propagates through the audio pipeline: TTS synthesis is aborted with a 100ms protection window to prevent abrupt cutoffs, active audio playback queues are cleared, and the speech processing pipeline transitions to listening mode. This multi-level architecture supports natural barge-in conversation patterns, allowing users to interrupt the agent's explanation at any point and resume seamlessly after their input is processed.

\begin{figure*}[t]
    \centering
    \includegraphics[width=0.99\linewidth]{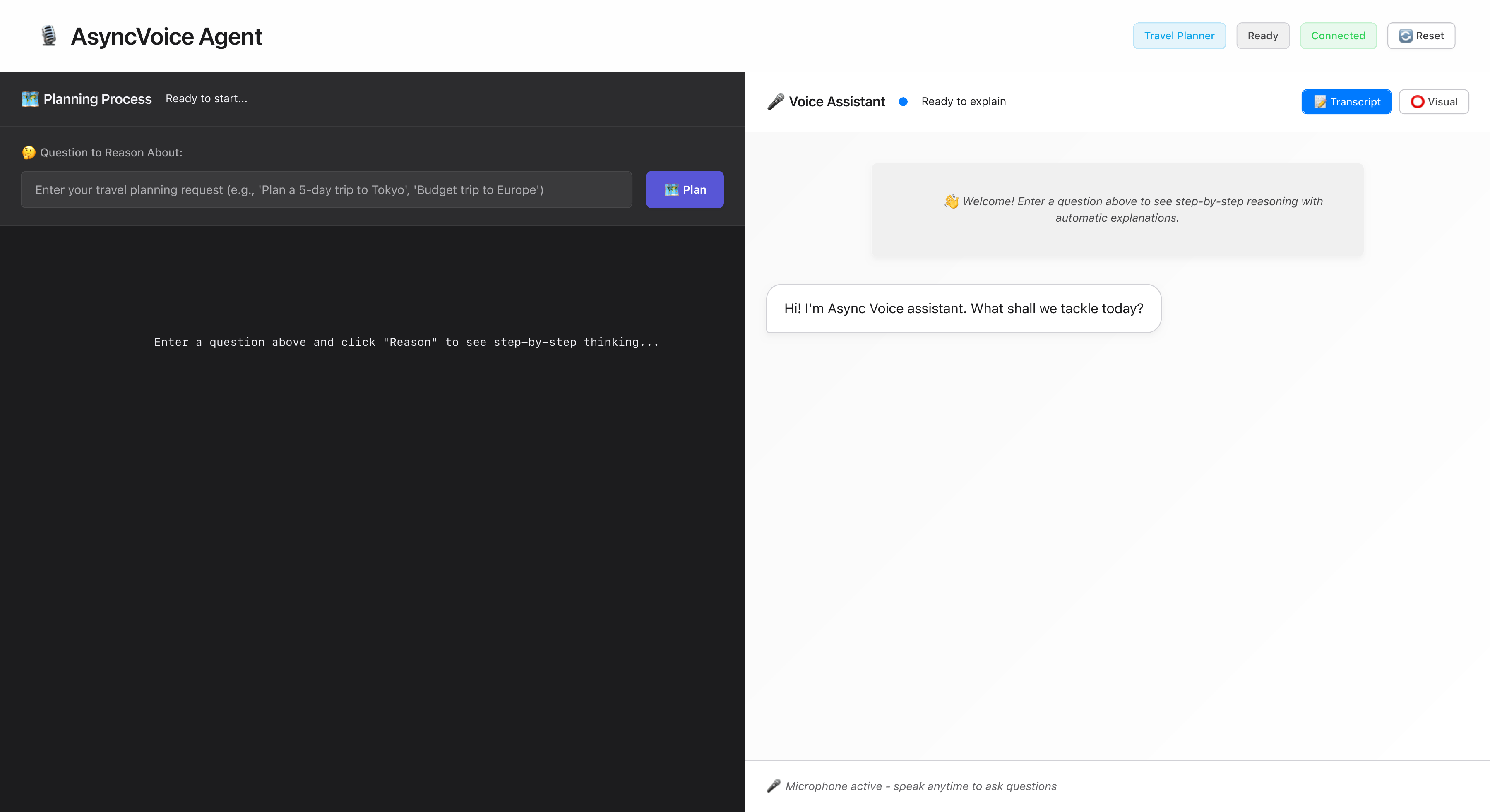}
    \caption{The front-end web interface for the AsyncVoice Agent. The user can enter a query to begin a task. The interface displays the live \texttt{Planning Process} and indicates that the microphone is always active for interruptions.}
    \label{fig:frontend}
\end{figure*}

\subsection{Frontend Real-Time Audio Processing}
The web client, whose interface is depicted in Figure~\ref{fig:frontend}, implements a modern audio processing pipeline using the Web Audio API with AudioWorklet processors. One processor handles microphone input, while another manages the audio output queue with interruption support. The interface provides the user with a real-time transcription display and a visualization of the CoT reasoning in the \texttt{Planning Process} panel. State management covers WebSocket connection status, audio session states (e.g., listening, processing), and progress tracking, ensuring the user always has clear visual feedback on the system's status.

%-------------------------------------------------------------------------
\section{Evaluation}
\label{sec:eval}
To objectively validate the benefits of our proposed architecture, we present a rigorous evaluation framework designed to quantify the performance of the AsyncVoice Agent. To ensure objective and reproducible results, the evaluation is conducted using an automated framework. This approach focuses on quantifiable metrics: responsiveness, reasoning quality, and process fidelity. We benchmark our system against two baselines across three distinct, pre-defined operational scenarios.

\subsection{Experimental Setup}
Our evaluation compares the performance of the following three systems:
\begin{itemize}
    \item \textbf{AsyncVoice Agent (Proposed)}: Our full system, featuring a backend MCP Server for reasoning and a GPT-4o explainer model for real-time verbalization.
    \item \textbf{Monolithic Agent (Baseline 1)}: A system that uses the same backend MCP Server but verbalizes the entire reasoning chain only after its complete generation.
    \item \textbf{Explainer-Only Agent (Baseline 2)}: A standard conversational system using GPT-4o for both reasoning and verbalization, without a decoupled backend.
\end{itemize}

\subsection{Evaluation Scenarios and Metrics}
We evaluate across three diverse reasoning scenarios using 100 queries for each: Math Solver problems are drawn from the GSM8K benchmark \cite{cobbe2021training} and require multi-step arithmetic; for Travel Planner and Deep Research, we used GPT-4o to generate scenarios with explicit constraints, such as budget limits for travel or minimum citation counts for research, that demand multi-round planning and analysis.

Our evaluation assesses performance using three core metrics. The first is \textbf{responsiveness}, measured by Time to First Audio (TTFA), which is the latency from user query submission to the start of audible output. The second is \textbf{reasoning quality}, assessed through hybrid frameworks combining automated scoring with GPT-4o assessment. For the Math Solver, this blends exact numerical accuracy (70\%) with reasoning methodology evaluation (30\%), while for the Travel Planner and Deep Research scenarios, it integrates constraint satisfaction with solution quality assessment. The third metric, \textbf{process fidelity}, is applied exclusively to our AsyncVoice Agent; here, GPT-4o scores the semantic and logical consistency between streamed explanations and the backend reasoning texts on a 1-to-5 scale. To ensure a fair comparison, complete reasoning traces for all systems are included in the evaluation.
\subsection{Results}
As summarized in Table~\ref{tab:evaluation_summary}, the AsyncVoice Agent demonstrates fundamental advantages in responsiveness, achieving TTFA of approximately 15ms across all scenarios, a 600-1,800$\times$ reduction compared to baseline approaches that enables genuine real-time interaction. While baselines occasionally achieve higher reasoning quality scores, the AsyncVoice Agent remains competitive. It is worth noting that for this evaluation, the backend MCP servers were configured for a single-pass reasoning process; enabling their native multi-round capabilities would likely improve reasoning scores further, but our focus is on interaction fidelity, not task-specific score optimization. This suggests that the streaming benefits do not significantly compromise reasoning accuracy. Critically, the high Process Fidelity scores validate that real-time explanations faithfully represent backend reasoning processes, addressing essential requirements for trustworthy human-AI collaboration.

\begin{table}[ht]
\centering
\caption{Objective evaluation results summary}
\label{tab:evaluation_summary}
\resizebox{\columnwidth}{!}{%
\begin{tabular}{@{}llccc@{}}
\toprule
\textbf{Scenario} & \textbf{Metric} & \textbf{Monolithic} & \textbf{Explainer-Only} & \textbf{AsyncVoice} \\ \midrule
\multirow{2}{*}{\textbf{Math Solver}} & TTFA (s) & $9.480$ & $4.618$ & \textbf{0.015} \\
 & Score & $96.36$ & $90.60$ & 92.20 \\ \cmidrule(l){2-5}
\multirow{2}{*}{\textbf{Travel Planner}} & TTFA (s) & $26.907$ & $12.089$ & \textbf{0.015} \\
 & Score & $96.40$ & $96.70$ & 91.80 \\ \cmidrule(l){2-5}
\multirow{2}{*}{\textbf{Deep Research}} & TTFA (s) & $27.184$ & $13.980$ & \textbf{0.014} \\
 & Score & $81.80$ & $72.60$ & 79.50 \\ \midrule
\textbf{All Scenarios} & Process Fidelity & N/A & N/A & \textbf{4.73} \\ \bottomrule
\end{tabular}%
}
\vspace{-5mm}
\end{table}

\subsection{Discussion and Limitations}
The performance gap versus monolithic baselines results from our one-way streaming design where explainability-optimized outputs and token limits constrain reasoning depth. We consider this an acceptable trade-off for achieving 600-1800× latency reduction. Our automated evaluation ensures reproducibility (using TTS-generated queries for audio consistency), while the system fully supports natural voice input with sub-100ms interruption handling and configurable explanation styles. Future work will propagate user feedback to upstream MCP servers, enabling human-in-the-loop refinement that could 
surpass static approaches. Current limitations include TTS prosody and unidirectional reasoning flow; however, these represent engineering challenges rather than fundamental architectural barriers.

%-------------------------------------------------------------------------
\section{Conclusion}
\label{sec:conclusion}
We introduced AsyncVoice Agent, the first system to enable real-time, interruptible dialogue with an LLM's live reasoning stream. By decoupling the reasoning backend from the explanation interface through asynchronous architecture, we transform passive CoT consumption into active collaboration. This work establishes a new paradigm for human-AI interaction in complex reasoning tasks, demonstrating that sub-second responsiveness fundamentally changes how users engage with AI reasoning processes. We believe this approach will inspire further research into interactive AI systems where human expertise guides model computation in real-time.

%-------------------------------------------------------------------------
% References
%-------------------------------------------------------------------------
{\small
\bibliographystyle{IEEEtran}
\bibliography{reference}
}

\end{document}